\documentclass{aa}
\usepackage{natbib}
\usepackage{graphicx,txfonts}
\usepackage{hyperref}

\bibpunct{(}{)}{;}{a}{}{,}

\begin{document}
	
	\title{Constraints on the progenitor models of fast radio bursts from population synthesis with the first CHIME/FRB catalog}
	
	\author{Min Meng\inst{1,2}
	\and  Can-Min Deng\inst{1}\fnmsep\thanks{Corresponding author: dengcm@gxu.edu.cn}}
	
	\institute{Guangxi Key Laboratory for Relativistic Astrophysics, Department of Physics, Guangxi University, Nanning 530004, China\\
	\and 
	Department of Physics, Faculty of Arts and Sciences, Beijing Normal University, Zhuhai 519087, China\\ }
	
    \date{Accepted April 23, 2025}
	
	\authorrunning{ Meng \& Deng}
	\titlerunning{Constraints on the progenitor models of FRBs}
	
	\abstract
	{Fast radio bursts (FRBs) are enigmatic extragalactic radio transients with unknown origins. We performed comprehensive Monte Carlo simulations based on the first CHIME/FRB catalog to test whether the FRB population tracks the cosmic star formation history directly or requires a delay. By fully considering CHIME’s complex selection effects and beam response, we find that the hypothesis that the FRB population tracks the SFH is not ruled out by the current data, although a small delay is preferred. This is consistent with the scenario in which young magnetars formed through core-collapse supernovae serve as the progenitors of FRBs. 	However, we estimate the local volumetric rate of FRB sources with energy above \(10^{38}\) erg to be \(2.3^{+2.4}_{-1.2} \times 10^5~\mathrm{Gpc}^{-3}~\mathrm{yr}^{-1}\), which is consistent with previous results. This high volumetric rate means the core-collapse magnetar scenario alone cannot fully account for the observed population. Further theoretical efforts are required to explore alternative or additional progenitor channels for FRBs.}

	\maketitle
	
	\section{Introduction}
	{Since the first discovery in 2007 \citep{2007Sci...318..777L}, hundreds of fast radio bursts (FRBs) have been observed. However, the physical origins of FRBs remain  unknown \citep{2019ARA&A..57..417C,2019A&ARv..27....4P,2020Natur.587...45Z,2021SCPMA..6449501X,2023RvMP...95c5005Z}. Many progenitor models have been proposed to explain the origin of FRBs \citep{2019PhR...821....1P}. The most popular is the magnetar model \citep{2014MNRAS.442L...9L,2017ApJ...843L..26B,2019MNRAS.485.4091M,2020ApJ...898L..29M,2020MNRAS.498.1397L}, which has been partially confirmed by the discovery of FRB 20200428 \citep{2020Natur.587...54C,2020Natur.587...59B}.  FRB 20200428 is an FRB-like radio burst from a young magnetar in the Milky Way(WM),  SGR 1935+2154 \citep{2020ApJ...898L..29M,2021NatAs...5..378L,2021NatAs...5..372R}, and this magnetar is associated with the supernova remnant G57.2+0.8 \citep{2018ApJ...852...54K,2020ApJ...898L...5Z}. In addition, observations of some FRB host galaxies show that they are located in a star-forming region \citep{2017ApJ...834L...7T,2020Natur.577..190M,2022Natur.606..873N}, which  leads to the speculation that  FRBs are produced by young magnetars born after the death of massive stars. However, the types of host galaxies are diverse, and some FRBs  are not located in star-forming galaxies or regions \citep{2020ApJ...903..152H,2022AJ....163...69B}. 
	A typical example is FRB 20200120E, located within a globular cluster in M81 \citep{2022Natur.602..585K}.
	This implies that FRBs likely also originate from old stellar populations, such as compact binary systems \citep{2016ApJ...823L..28G,2020MNRAS.497.1543G,2020ApJ...890L..24Z,2020ApJ...893L..26I}. Furthermore, although FRB 20180916B and {FRB 20121102A} are located in star-forming regions, their long periodicity indicates a possible association with binary systems \citep{2020Natur.582..351C,2020MNRAS.495.3551R,2021ApJ...922...98D,2021ApJ...917...13S,2021ApJ...918L...5L}.
	The discovery of FRB 20200120E highlights the diversity and complexity of  FRB origins.}

	{Faced with such a variety of progenitor models and diverse observational results, population studies, particularly those analyzing the volumetric rate and its cosmic evolution (i.e., redshift distribution), would provide valuable insights into FRB origins. This is because different progenitor models predict distinct population properties of FRBs. 
	In this regard, many authors have studied the volumetric rate and the redshift distribution of FRBs based on the early, small samples from Parkes and ASKAP
	\citep{2019JHEAp..23....1D,2019A&A...625A.109L,2019A&A...632A.125G,2020MNRAS.498.3927H,2021MNRAS.501.5319A,2021A&A...651A..63G,2021A&A...647A..30G,2021MNRAS.501..157Z,2022MNRAS.510L..18J}.
	However, due to the small sample size, it was difficult for those early studies to give effective constraints on the population properties of FRBs. Thanks to CHIME's large field of view and sensitivity, hundreds of FRBs have been detected so far \citep{2021ApJS..257...59C}.  Studies on the energy function and redshift distribution of FRBs based on this large and uniform sample have been carried out.}
	
	{ \cite{2022ApJ...924L..14Z}, \cite{2022JCAP...01..040Q}, and \cite{2023A&A...675A..66Z} performed Monte Carlo simulations to directly test the hypothetical distribution of the FRB redshifts, and their results ruled out the assumption that the FRB redshift distribution simply follows the star formation history (SFH). However, if there is relatively long delay with respect to the SFH, the simulation results can be consistent with the observational data. We note that, perhaps due to the computational limitations in their Monte Carlo simulations, they only conducted simulations with a limited set of parameter combinations, potentially leading to results that such simulations may not be comprehensive.
	\cite{2024ApJ...962...73L} and \cite{2024ApJ...969..123L} constructed  a Bayesian framework to fit the observed number of FRBs within specific parameter spaces to constrain their redshift distribution model. Again, they ruled out the hypothesis that the FRB population simply traces the SFH, and a delay of several billion years with respect to the SFH is preferred by their data. The nonparametric methods commonly used in other fields have also been used to study this problem.
	By using the $V_{\rm{max}}$ method, \cite{2022MNRAS.511.1961H} also ruled out the SFH scenario for the FRB redshift model.
	\cite{2024ApJ...973L..54C} and \cite{2024arXiv240600476Z} use the Lynden-Bell c$^{-}$ method to directly derive the volumetric rate and the redshift distribution of FRBs without assumptions. They obtain a  volumetric rate  at redshift $z=0$ , which then decays rapidly with a single power law with $z$. In other words, their results also show that the FRB redshift distribution does not follow the SFH.	}
	
     {However, it is important to note that the studies mentioned in the previous paragraph all have significant shortcomings. Firstly, due to insufficient localization accuracy, the fluence measurements provided in the first CHIME/FRB catalog have not been corrected for the beam response. As a result, the fluences listed in the catalog are not the true values but merely lower limits \citep{2021ApJS..257...59C}. The CHIME/FRB team recently released beam-corrected fluences for a subset of FRBs, showing that most of the corrected values are significantly higher than the lower limits reported in the catalog \citep{2024ApJ...969..145C}. Therefore, directly using these lower limits as true fluences in studies would clearly lead to questionable results. Secondly, the beam pattern of CHIME/FRB is very complex, which in turn results in equally complex selection effects. In addition, CHIME/FRB selects strongly against events with low dispersion measures (DMs) due to radio frequency interference (RFI) filters \citep{2021ApJS..257...59C}. Thus, a proper correction for selection effects must take these complexities into account, and the methods used to handle selection effects in those studies are overly simplistic.}
	
	{\cite{2023ApJ...944..105S} comprehensively took selection effects into account and used the signal-to-noise ratio (S/N) as a proxy of fluence, with the help of CHIME's injection system. Then they constructed a Poisson likelihood within the  Bayesian framework to fit the number of FRBs within a certain parameter range, attempting to constrain the redshift distribution models of FRBs. While their study provided valuable insights, it did not sufficiently constrain the redshift distribution model of FRBs. This highlights the importance of exploring alternative methods that can effectively constrain the redshift distribution models of FRBs, which is a key factor in understanding their origins.  }
	
	{In this work we test the redshift distribution models of FRBs using Monte Carlo simulations that can model the instrumental selection effects in detail. 
	We also simulated the beam response based on the beam model from \cite{2023AJ....165..152M} so that we could appropriately use the fluence (lower limit) given in the first  CHIME/FRB catalog.
	We  developed a parallel-computable code hat runs on a thread-ripper CPU; this helped us  systematically and thoroughly explore the potential parameter space of the models, ensuring that no parameter space that could possibly match the data is not omitted. From this, we wished to determine  whether  the  data can tell us something about the redshift distribution of FRBs and, in particular, whether the SFH model can be ruled out.}
	
	\section{Simulations} 
	{The first CHIME/FRB catalog contains 474 non-repeating sources and 62 bursts \footnote{ {To minimize the interference of the repetitiveness of repeating sources on the results during the analysis process, we adopted the treatment method of only selecting the first burst of repeating sources.As noted by \cite{2021A&A...647A..30G, 2023PASA...40...57J}, this method still introduces selection biases, leading to an underrepresentation of FRBs in the nearby Universe. Despite this limitation, it remains a widely adopted approach in FRB population studies. The bias introduced by repetition is expected to be minimal after applying this method, since repeat bursts are well distributed within the current sample.}}
	from 18 repeating sources \citep{2021ApJS..257...59C}.  
	However, this full sample suffers from selection effects that are difficult to account for, making it necessary to apply appropriate filtering to minimize their impact. Based on this consideration, \cite{2023ApJ...944..105S} applied several selection criteria to create their sample, finding 225 bursts in the catalog.
	{In this work, we adopted their refined sample but excluded two bursts (FRB 20190307A and FRB 20190531B) due to their zero fluence values \citep{2021ApJS..257...59C}. These anomalies were caused by system restarts during data collection, which prevented an accurate beamformer calibration due to the lack of steady-source transits prior to upstream complex gain calibration. As a result, the fluence could not be properly measured. Therefore, our final analysis is based on a sample of 223 bursts \footnote{https://github.com/kaitshin/CHIMEFRB-Cat1-Energy-Dist-Distrs/blob/main/data/cat1\_sample.dat}.  In the subsequent part of this paper, any reference to the FRB sample specifically refers to this dataset.}
	
	{The fundamental essence of Monte Carlo simulations is to use hypothetical models to simulate several observable quantities while considering instrumental selection effects. Then, the simulated sample is compared with the observed  one to determine whether the assumed models are rejected.  In the case of FRBs, the primary observable quantities are the specific fluence ($ F_{\nu}$), and the DM\footnote{We employed the DMs designated as "$ \rm bonsai\_dm$" and "$\rm dm\_exc\_ne2001$", as well as the fluence identified as "$\rm fluence $" in the first CHIME/FRB catalog.}. Therefore, our basic procedure was to simulate the fluence and DM of the FRBs and then perform a Kolmogorov-Smirnov (KS) test by comparing them with the observed sample. }

\subsection{Redshift distribution model} 
	\medskip\noindent\textnormal{\textit{SFH model}.\,}   
	{As discussed in the introduction, the currently popular FRB models speculate that  FRBs originate from the activity of compact stars. The compact stars result from the evolution of stars, and hence the redshift distribution of FRBs may be connected to the SFH in some way. Therefore, we considered two redshift distribution models for FRBs, both of which are relevant to the SFH.}
	
	{In the simplest case, the formation rate density of FRBs directly follows the SFH. In this case, the formation rate density of FRBs  evolves  with redshift ($z$) as}
	
	\begin{equation}
		\frac{ dN}{ dtdV} \propto{\rm {SFH}}(z)= [(1+ z)^{a\eta}+(\frac{1+ z}{ B})^{ b\eta}+(\frac{1+ z}{ C})^{ c\eta}]^{\frac{1}{\eta}},
		\label{sffr}
	\end{equation}
	{where $ a=3.4,  b=-0.3, c=-3.5,\eta=-10$, $B=5000 $,  and $C=9 $ are adopted \citep{2008ApJ...683L...5Y}.}
	
      \medskip\noindent\textnormal{\textit{Delayed SFH model}.}
	{Based on the fact that many host galaxies of
	FRBs are late-type galaxies, and FRB 20200120E being located within a globular cluster, there is speculation that FRBs might also originate from old stellar systems \citep{2020ApJ...903..152H,2022AJ....163...69B,2022Natur.602..585K}. In such a scenario, the formation rate of FRBs would exhibit a certain time delay  with respect to SFH, which is given by}
	
	\begin{equation}
		\frac{dN}{dtdV} \propto \int_{z}^{\infty}{\rm {SFH}}(z^{'})f[ t(z^{'})-t(z)]\frac{ dt}{ dz^{'}} dz^{'},
		\label{com}
	\end{equation}  
	{where $f(\tau)$ is the time delay distribution, }
	
	\begin{equation}
		f(\tau)=\frac{1}{\tau \sigma_{\tau}\sqrt{2\pi}} {\rm{exp}}[-\frac{( \rm {ln}\tau- \rm {ln} \bar{\tau})^2}{2{ \sigma}^2_{\tau}}], ~\tau>0
		\label{tt}
	\end{equation} 
	{Here  $\bar{\tau}$\ is the mean of time delay, which is a free parameter in the simulations with a  prior range of  [0, 10] Gyr, and $ \sigma_{\tau}$ = 0.8 is adopted \citep{2015MNRAS.448.3026W,2022ApJ...924L..14Z}.  For a given redshift, the corresponding look back time is}
	
	\begin{equation}
		t=\int_{0}^{z} \frac{ dz^{'} }{(1+z^{'}) H(z^{'})}  ,
		\label{tz}
	\end{equation} 
	{where $H(z)={ H_0} \sqrt{\Omega_m \left(1+z\right)^3+\Omega_\Lambda}$ is the Hubble parameter, and $ H_0$, $ \Omega_m$, $ \Omega_\varLambda$ are cosmological constants whose values are adopted from the Planck results \citep{2020A&A...641A...6P}. }

	{Once the distribution of the formation rate density is determined, the observed redshift distribution of FRBs can be  derived as \citep{2021MNRAS.501..157Z}}
	
	\begin{equation}
		\frac{ dN}{ dt_{\rm obs}  dz}=\frac{1}{1+ z} \frac{ dN}{ dt dV} \frac{ dV}{ dz}.
		\label{dndtdz}
	\end{equation}
	{The comoving volume per unit redshift can be described as}
	
	\begin{equation}
		\frac{ dV}{ dz} = \frac{4 \pi c D_L^2}{(1+ z)^2  H(z)},
		\label{dvdz}
	\end{equation} 
	{where $D_L$ is the luminosity distance.}

	{The spectral index for an individual FRB describes the change of the radiation intensity with frequency. Based on the FRBs detected by ASKAP, \cite{2019ApJ...872L..19M} report  an average spectral index of $\gamma \approx -1.5$. In this scenario, all FRBs are assumed to be broadband, and each observed FRB has a consistent spectral index.  
    However, \cite{2021ApJ...923....1P} found that FRBs are likely to be narrowband. This finding leads us to adopt the rate interpretation when modeling the spectral properties of FRBs. 
    The factor of $\left(1+z\right)^{\gamma}$ is incorporated into the source rate evolution models as the rate interpretation of the spectral index. Then the rate density of FRB can be calculated as }
	
	\begin{equation}
		\frac{ dN}{ dt_{\rm obs}  dz}=\left(1+z\right)^{\gamma}\frac{1}{1+ z} \frac{ dN}{ dt dV} \frac{ dV}{ dz}.
		\label{dndtdzz}
	\end{equation} 	
	{ In the rate interpretation, each FRB is observed according to the sensitivity at its specific frequency. For instruments with widely spaced beams, the observed sky area scales with frequency, meaning that low-frequency bursts are preferentially detected, even for a frequency-independent rate. This introduces an observational bias, favoring detection of  those FRBs with steeper spectral indices. An accurate modeling approach would involve correcting for these biases to better estimate the true distribution of FRBs across the spectrum. Indeed, a strong correlation between spectral properties and inferred source evolution has been observed by both  \cite{2023ApJ...944..105S}  and \cite{2025PASA...42...17H}. 
   \cite{2022MNRAS.509.4775J} suggested that making a correction of 0.85 to $\gamma$ can account for these biases. Therefore,  we adopted  $\gamma=-0.65$ in this study \citep{2022MNRAS.509.4775J}.\footnote{Although the sample we used was from CHIME/FRB, the results of \cite{2023ApJ...944..105S}  show that the average spectral index of CHIMIE FRBs may be similar to that of the ASKAP FRBs.} \footnote{To justify our choice of fixing \(\gamma = -0.65\) in our analysis, we tested the impact of spectral variation by running simulations with \(\gamma = -0.65\), \(-1.0\), and \(-1.5\). The results showed only minor differences, indicating that our conclusions remain robust against moderate changes in \(\gamma\). While, in principle, \(\gamma\) could be treated as a free parameter, doing so would significantly increase the computational load and extend the simulation time substantially. Given this trade-off, we adopted \(\gamma = -0.65\) as a representative value to balance accuracy and computational feasibility.}
	}

	\subsection{Dispersion measure} 
	{Observationally, the DM value of each FRB can be measured. And theoretically, the DM value of an FRB consists of the following components:}
	
	\begin{equation}
		\rm DM=\rm DM_{\rm WM} + \rm DM_{\rm halo}+\rm DM_{IGM} +\frac{\rm DM_{host}}{1+\it {z}},
		\label{eq:dme+} 
	\end{equation}
	{where the MW contribution ($\rm DM_{\rm WM} $) is derived from the MW electron density model NE2001 \citep{2002astro.ph..7156C}. Following \cite{2022ApJ...924L..14Z}, a fixed value for the MW halo contribution $\rm DM_{\rm halo}=30~ pc ~ cm^{-3}$ is adopted. The cosmological redshift factor ($1 + \it z$) is used to convert the DM observed by a stationary observer to an observer on Earth \citep{2014ApJ...783...35T}. }

	{When a redshift z is randomly drawn from Eq.(\ref{dndtdz}), one can calculate the mean value of the intergalactic medium contribution $\overline{ \rm DM}{\rm _{IGM}}$  as  }
	
	\begin{equation}
		\overline {\rm DM}{\rm _{IGM}}(z) = \frac{ 3c H_0^2 \Omega_b f_{\rm {IGM}} f_e }{8\pi G_r m_p} \int_0^z \frac{\left(1+z\right)^{'} dz^{'}} {H(z^{'})},
		\label{eq:mcq} 
	\end{equation}
	{where $\Omega_b$ a cosmological parameter whose value is adopted from the Planck results \citep{2020A&A...641A...6P}, $G_r$ is the gravitational constant, $m_p$ is proton mass, $f_e=7/8$ is the free electron number per baryon in the Universe, and $f_{\rm IGM}=0.84$ is the fraction of baryons in the intergalactic medium. }
	
	{The electron number density along different sight lights is not uniform \citep{2014ApJ...780L..33M}, and the exact value of  $\rm DM_{IGM}$ for a given $z$ can be described by the following distribution \citep{2020Natur.581..391M},}

	\begin{equation}
		f \rm_{IGM}(\bigtriangleup) \propto \bigtriangleup^{-\beta}~exp\left(-\dfrac{\bigtriangleup^{-\alpha}-{\emph C_0}}{2\alpha^{2}\sigma_{IGM}^{2}}\right), 
		\label{eq:fIGM} 
	\end{equation}
	{where  $\rm \bigtriangleup =DM_{IGM}/~\overline{ \rm DM}{\rm _{IGM}}$, $\rm \sigma_{DM}$ is the effective standard deviation, $C_0$  is chosen such that the mean of the distribution is unity, $\alpha$ and $\beta$ are two indices related to the inner density profile of gas in halos. We adopted the best fit values of $C_0$ and $\rm \sigma_{IGM}$ given by \cite{2021ApJ...906...49Z} and the indices $\alpha=3$ and $\beta=3$.}
	
	{Since  little is known about the distribution of $\rm DM_{\rm host}$, we simply modeled it using the lognormal distribution}
	
	\begin{equation}
		\begin{split}
			%\begin{align*}
			f(\rm DM_{host})=\frac{1}{\rm DM_{host} ~\sigma_{host}\sqrt{2\pi}} \hspace{3cm} \\
			\times \rm exp[-\frac{\left( \rm {\log} ~\rm DM_{host}- \rm {\log} ~{\rm\overline {DM}_{host}}\right)^2 }{2{ \sigma}^2_{host}}],
			\label{dm}
			%\end{align*}
		\end{split}
	\end{equation}
	{where  $\log{\rm\overline {DM}_{host}}$ is the mathematical expectation of the distribution and is a free parameter; in the simulations with a prior range of $[10,1000]~ \rm pc~cm^{-3}$, a fixed value of $\sigma_{\rm host}=0.4$ was used \footnote{When we let  $\sigma_{\rm host}$  as a free parameter in the simulation, it is not constrained by the data at all. Its value has little impact on the results when it is within the range of 0.2 to 0.8. Therefore, we chose a fixed value of 0.4 for it.}.  }
	
	{Because the contribution from the MW is relatively easy to subtract, people commonly combine the extragalactic contribution to DM and define it as }
	
	\begin{equation}
		\rm DM_E=\rm DM- \rm DM_{\rm WM} + \rm DM_{\rm halo}=\rm DM_{IGM} +\frac{\rm DM_{host}}{1+\it {z}},
		\label{eq:dme} 
	\end{equation}
	{which can be directly compared with the model values. Following \cite{2023AJ....165..152M}, we cut the case of $\rm DM_E >5000 ~pc~cm^{-3}$ directly in the simulation.
	}
	
	\subsection{Energy distribution } \label{energy Distribution }
	{For an FRB originating in the Universe, whether it can be observed is not only related to its redshift but also to its energy. To model the energy of FRBs, the cut-off power-law function \citep{1976ApJ...203..297S} is commonly used in the literature}
	
	\begin{equation}
		\frac{ dN}{ dE}\propto{ \left(\frac{ E}{ E_c}\right)}^{\alpha}e^{-\frac{ E}{ E_c}},
		\label{eq:dNdE} 
	\end{equation}  
	{where $\rm \alpha$ is the power-law index, and ${ E_c}$ is the cutoff energy.{ We analyzed the energy distribution of FRBs of the first catalog of CHIME/FRB. In the simulation, the energy of FRBs will be generated in the range of $[10^{38}, 10^{43}]$, based on the fact that  the energy of the majority of the sources is \(\gtrsim 10^{38}\) erg \citep{2023ChPhC..47h5105T}.
	And the  prior range of $\alpha$ is taken as $[-3,-1]$, and of  $ E_c$ is taken as $[10^{41.5},10^{43}]$ erg.}

	{For a mock FRB with its isotropic energy $E$ and redshift $z$, the specific fluence received by the observer can be calculated as \citep{2018ApJ...867L..21Z} }
	\begin{equation}
		\mathcal{F}_{\nu}= \frac{\left(1+\it z\right) E }{ 4\pi D_L^2 \nu_c},
		\label{eq:Fu} 
	\end{equation}
	{where $\nu_c= $ 600 MHz is the central observing frequency of CHIME/FRB. \cite{2023AJ....165..152M} found that, for CHIME FRBs,  the fluence is no longer a reliable indicator of the S/N when it exceeds 100 Jy ms due to RFI  rejection. This leads to a bias against high-fluence FRBs. To reduce the impact of this selection effect in our simulated FRB population, we limited our simulations to bursts with fluences below 100 Jy ms, which ensures a more reliable comparison with the observed data.}

	\subsection{ Instrument selection effects  }
	{In practice, the ultimate criterion for determining the detection of an FRB is the signal-to-noise ratio (S/N), which can be calculated as \citep{2023AJ....166..138A}}
	
	\begin{equation}
		{\rm S/N} =\frac{G\mathcal{F}_{\nu}\sqrt{B    N_p}}{\left(T_{\rm rec }+T_{\rm sky }\right) \sqrt{w} }\times{s^{2/3}(\mathrm{DM})},
		\label{eq:S/N} 
	\end{equation} 
	{where $G$ is the system gain in $\rm K/Jy,$ which is determined by the beam response of CHIME/FRB (we discussed this in more detail in the next paragraph\footnote{Here, in order to maintain the simplistic conversion of energy to S/N for the CHIME/FRB telescope, we absorbed the complexity of the beam response into the beam-corrected gain ($G$).}), $\mathcal{F}_{\nu}$ is the fluence of the signal in $\rm Jy~ms$, $T_{\rm rec }$ and $T_{\rm sky }$ are the receiver and the sky temperature in $\rm K,$ respectively, $B$ is the bandwidth in kHz (which is 400 MHz for CHIME/FRB), $w$ is the observed pulse width in ms, and $ N_p=2$ is the number of polarizations. For the receiver temperature, we adopted its typical value, $T_{\rm rec }=50$ K \citep{2018ApJ...863...48C}. The sky temperatures is estimated from the Haslam sky survey \citep{1982A&AS...47....1H}, where the survey frequency is scaled to the CHIME’s central frequency by adopting a spectral index of -2.6 for the Galactic emission. Then $T_{\rm sky }$ is randomly generated in the range of [7K, 12K], which  roughly corresponds to the temperature range of the celestial region scanned by CHIME/FRB at 600 MHz. And it should be realized that a small uncertainty on $T_{\rm sky }$ has little impact on our results, because $T_{\rm rec }+T_{\rm sky }$ is as large as  $\sim$ 60K. Finally, a detection threshold of $\rm S/N=12$ is applied for the CHIME/FRB sample.}

	{\medskip\noindent\textnormal{ \rm \textit{DM selection}.}}
	{As noted in \cite{2021ApJS..257...59C}, the CHIME/FRB telescope exhibits strong selection effects for FRBs with low DM, primarily due to RFI filtering. This introduces significant challenges in accurately modeling the observed FRB population, particularly at low DM. To address this, we incorporated the DM selection function, \( s(\mathrm{DM}) \), into our modeling to fully account for its impact.  
	The \( s(\mathrm{DM}) \) represents the relative fraction of FRBs retained after selection cuts as a function of DM. Following \cite{2023PASA...40...57J}, we fit this function using a fourth-order polynomial, and the peak value of the selection function is normalized to 1. The parameters of the maximum likelihood fit are provided in table \ref{tab:fun}. To incorporate the DM selection effects into the S/N, we applied a Euclidean scaling relationship between the event number and the S/N, as described in \cite{2023PASA...40...57J}, such that  \( \mathrm{S/N_{bias}} \sim s^{2/3}(\mathrm{DM}) \). This correction has been applied to the S/N in equation (\ref{eq:S/N}), ensuring that the modeled S/N reflects the DM-dependent selection biases.  
	}
	
	\begin{table}
		\centering
		\caption{ Valid range of free parameters for the SFH model. }
		\renewcommand{\arraystretch}{1.25} 
		\resizebox{0.5\textwidth}{!}{ 
			\begin{tabular}{ccccccc}
				\hline
				\hline
				Parameters  &$N$&$a4$&$a3$&$a2$&$a1$&$a0$ \\ 
				Value &$0.90$&$0.22$&$-2.55$&$10.25$&$-16.31$ &$8.99$\\
				\hline
			\end{tabular}
		}
		\label{tab:fun}
	\end{table}

	\medskip\noindent\textnormal{\textit{Beam response}.}
	{The beam pattern of the CHIME/FRB is very complex and is characterized by rapidly varying sensitivity across both the field of view and bandwidth. Recently, a preliminary version of CHIME/FRB beam model has been released\footnote{https://github.com/chime-frb-open-data/chime-frb-beam-model}  \citep{2023AJ....165..152M}. From the beam model, we obtained the beam-corrected gain ($G$) {at 600 MHz} used in equation (\ref{eq:S/N}), which is shown in figure \ref{fig:SEED}. Clearly, the gain is a function of the location $(x,y)$ in the topocentric coordinate system where $(x,y)=(0,0)$ is the zenith. With the zenith as the origin, $y$ is degrees north from the zenith and $x$ is degrees west from the meridian. In this work, we assumed that y ranges from $-60$ to 60 based on the fact that the y coordinates of the observed samples fall within this interval.}
	
	{Due to the limitations of localization for FRB sources, we recalculated the fluences that each burst was detected at beam boresight, where "boresight" means along the meridian of the primary beam (at the peak sensitivity of the burst decl.) \citep{2023AJ....166..138A}. Therefore, the  measured fluence, ${F_{\nu }}$, is not equal to its true value, ${\mathcal{F}_{\nu}}$, and the relationship between them is}
	
	\begin{equation}
		{F_{\nu }}=\frac{G(x,y)}{G(0,y)}{\mathcal{F}_{\nu}}.
		\label{eq:Fv}
	\end{equation} 
	{One sees that  the measured fluences ${F_{\nu }}$ {is always} less than or equal to the true values ${\mathcal{F}_{\nu}}$, because the gain $G(x,y) \leq G(0,y)$  is always true. Thus, the measured fluences are biased low with respect to the true values,  which corresponds exactly to the case of the fluences (lower limit) given in the first CHIME/FRB catalog \citep{2021ApJS..257...59C,2023AJ....166..138A}.  When performing the KS test, the measured fluences ${F_{\nu }}$ given by equation (\ref{eq:Fv}) are compared with the fluences (lower limit) of the catalog.}

	\begin{figure}
		\centering
		\includegraphics[width=1\hsize]{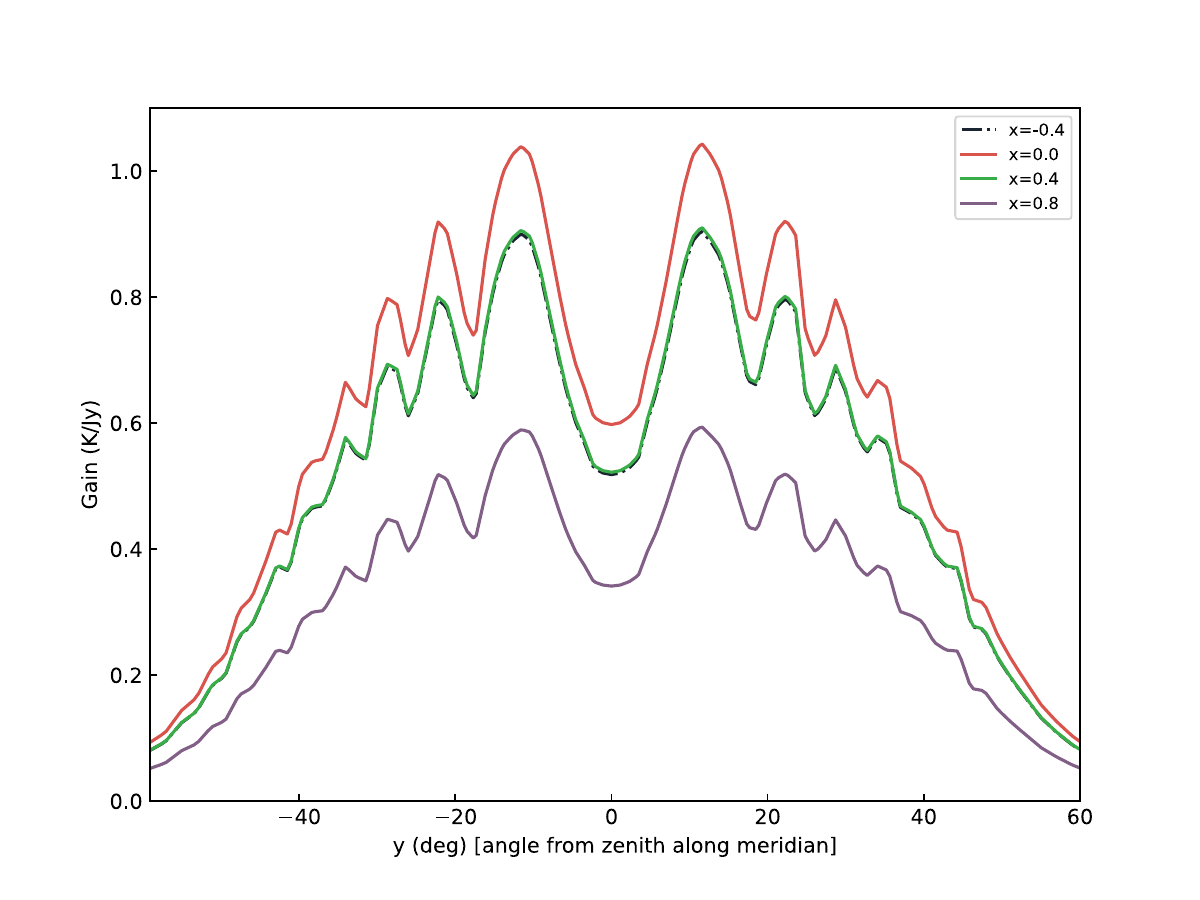}
		\caption{Beam-corrected gain ($G$) as a function of the angle from zenith along the meridian (y) at 600 MHz for $x=-0.4$, 0, 0.4, and 0.8. }
		\label{fig:SEED}
	\end{figure}
	
	\medskip\noindent\textnormal{\textit{Pulse width}.}
	{The observed pulse width of an FRB, {applied in equation \ref{eq:S/N}}, can be described as the sum of several components:}
	\begin{equation} 
		w{^{2}} = w{^{2}}_{\rm{int}} + w{^{2}}_{\rm{sc}} + w{^{2}}_{\rm{DM}} + w{^{2}}_{\rm{sample}}
	\end{equation}
	{where $w_{\rm{int}}$ is the intrinsic width, $w_{\rm{sc}}$ is the scattering time due to the propagation, \( w_{\rm{DM}}\) is the DM smearing at the telescope, and $w_{\rm{sample}}=0.983$ ms is the sampling time. { $w_{\mathrm{DM}}$ is calculated as $w_{\mathrm{DM}}=(8.3 \mu\text{s}) \mathrm{DM}\Delta\nu_{\mathrm{MHz}}\nu_{\mathrm{GHz}}^{-3}$, where $\Delta\nu= 24.4 $ kHz is FRB search frequency resolution and $\nu=600$ MHz is the central frequency of the telescope \citep{2018ApJ...863...48C}.  }The intrinsic width is not known, and there is no reliable formula for calculating $w\rm_{sc}$ \citep{2016ApJ...832..199X}.  The scattering time $w\rm_{sc}$ can be attributed to the following contributions, which are the MW part ($w\rm_{sc,MW}$), the intergalactic part ($w \rm_{sc,IGM}$), and the host galaxy part ($w\rm_{sc,host}$). The \( w_{\rm{sc,MW}} \) is described well by an empirical \( w_{\rm{sc}} - \rm DM \) relation \citep{2015ApJ...804...23K,2016arXiv160505890C}, but the relation for \( w_{\rm{sc,IGM}} \) is less certain. \cite{2017ApJ...835...29Y} derived a relation based on 17 FRBs, most of which lacked redshift measurements, but we find this relation tends to overestimate \( w_{\rm{sc,IGM}} \) for large \( \rm DM_{\rm{IGM}} \). We observe no clear \( w_{\rm{sc,IGM}} - \rm{DM} \) relation from a sample of several dozen FRBs with redshift measurements (in preparation). The understanding of \( w_{\rm{sc,host}} \) is even more limited. Given the difficulties in calculating both $w_{\rm{int}}$ and \( w_{\rm{sc}} \), we chose to model them according to the fiducial models derived by \citep{2021ApJS..257...59C}, see their Table 4, which may provide a more accurate representation of the intrinsic width and scattering width.}

	\subsection{Simulation procedures}\label{Simulations}

    {We assumed the distribution of FRBs in the sky is uniform and isotropic.} We performed the simulations as follows.
    
	 {Firstly, we randomly selected a set of parameter combinations from the corresponding model\footnote{The free parameters in the simulation for the corresponding model and their prior ranges are shown in table \ref{tab:sfr} and table \ref{tab:delay}.}.}
	
	 $\rm (i)$ We randomly selected a redshift \(z\) in the range $(0,8)$ according to Eq. (\ref{dndtdz}) and then sampled the corresponding $\rm DM_{{IGM}}$ for a given \(z\) based on the $f \rm_{IGM}(\vartriangle)$.
	
	 $\rm (ii)$ We randomly selected a ($ {\log} ~\rm DM_{host}$) in the range $(1,3)$ according to the $f (\rm DM_{host})$ and then determined the corresponding $\rm DM_{E}$ using Eq. (\ref{eq:dme}).
	
	$\rm (iii)$ We sampled the energy ($E$) in the range $(10^{37} ,10^{43})$ erg according to the energy distribution (\ref{eq:dNdE}) and used Eq. (\ref{eq:Fu}) to obtain the  $\mathcal{F}_{\nu}$ for the redshift (\(z\)) obtained in step $\rm (i)$.
	
	{$\rm (iv)$ We selected a $w$ randomly from the  observed distribution of  the pulse width and randomly selected a set of azimuth coordinates of $(x,y)$ in the ranges $-0.4 \leq x  \leq 0.8$ and $-60 \leq y  \leq 60$, accordingly obtaining a $G(x,y)$.  
	}
	
	{$\rm (v)$ We calculated the S/N of bursts according to equation (\ref{eq:S/N}).  If $\rm S/N \geq 12$, the burst is determined to be detectable. We measured its fluence using Eq. (\ref{eq:Fv}) and mark this burst as $( F_\nu,{\rm DM_{{E}}})_i$, where $i$ represents the $i$th burst. We repeated (i)-(v) until we got a mock sample containing 1000 FRBs, which is larger than the observed sample for a set of parameter combination \footnote{ In theory, simulating a larger number of bursts would enable a more rigorous calculation of the p-value and better capture the characteristics of the sample. However, limited computational power makes it challenging to simulate larger samples for each parameter set}.}
	
	 {By repeating this process, we were able obtain as many mock samples as we wanted, each simulated based on a set of randomly selected model parameters.}
	
	\section{Results  \label{sec:analysis}}
	{We compared the CHIME/FRB sample with the three model samples individually. The 1D KS tests were performed individually on the distributions of the specific fluence ($ F_{\nu}$) and the extragalactic DM ($\rm DM_E$). Considering that $ F_{\nu}$ and  $\rm DM_E$ are likely correlated \citep{2021ApJS..257...59C}, the 2D KS tests are performed to the 2D distribution of $( F_\nu,{\rm DM_{{E}})}$.  To reduce the error of the KS test, we  used the bootstrap method  to calculate the p-value following \cite{2021ApJS..257...59C}\footnote{https://github.com/syrte/ndtest}. }
	
	For the SFH model, the free parameters are $\alpha$ and $E_c$, which control the energy distribution of FRBs. Combinations of $\alpha$ and $E_c$  were randomly selected, and numerous mock samples were simulated, in which each mock sample contained 1000 FRBs. {We find that the SFH model cannot be rejected by the data  within a  considerable parameter space. We selected a set of parameter combinations from the simulations that yield the largest $p_{\rm 2DKS} $ and created a comparison plot between the observed sample and the simulated sample (see Fig. \ref{fig:comp1}). We selected 5000 parameter combinations with $p_{\rm 2DKS} \geqslant 0.01 $  and plotted their contour map in Fig. \ref{fig:sfr}.  The marginalized 1D distribution of each parameter is also plotted in histograms with $p_{\rm 2DKS} \geqslant 0.05$. It can be  seen from the figure that the data have a clear constraint on $\alpha$, and can only have an upper limit for $\overline{\rm DM}_{\rm host}$, while for ${\rm {log}}(E_c)$, there is no constraint, within the prior ranges, also presented in Table \ref{tab:sfr}.  Based on the marginalized 1D distribution of the parameters, one can infer a 95\% confidence level for the  parameters of  $ \overline{\rm DM}_{\rm host} <239 ~ \rm pc ~cm^{-3} $  and of $ \alpha=-2.1^{+0.5}_{-0.4} $, where the peak of the marginalized distribution is used as the central point\footnote{We caution that setting a p-value threshold and accepting only parameters within this range may result in an artificial uncertainty. In principle, if we were able to simulate a sufficiently large set of parameter combinations, this issue would be minimal. But due to computational limitations, we are currently unable to do that.  However,  we believe that this limitation does not significantly impact the conclusions of this paper.}.}

	\begin{table}
		\centering
		\caption{ Accepted range of free parameters for SFH model. }
		\renewcommand{\arraystretch}{1.3} 
		\begin{tabular}{cccccccccc}
			\hline\hline			
			Free parameters  & Prior & Accepted range\\
			\hline
			$\alpha$ & $[-3,-1]$& $[-2.5, -1.6]$ \\
			${\rm log}(E_{\rm c}/ \rm erg)$ &$ [41.5,43]$& [41.5, 43]\\		
			$\rm {\overline{\rm DM}_{host}/pc~ cm^{-3}} $& $[10,1000]$&[10, 239]\\
			\hline
		\end{tabular}
		\label{tab:sfr}
	\end{table}
	
		\begin{figure}
		\centering
		\includegraphics[width=1\hsize]{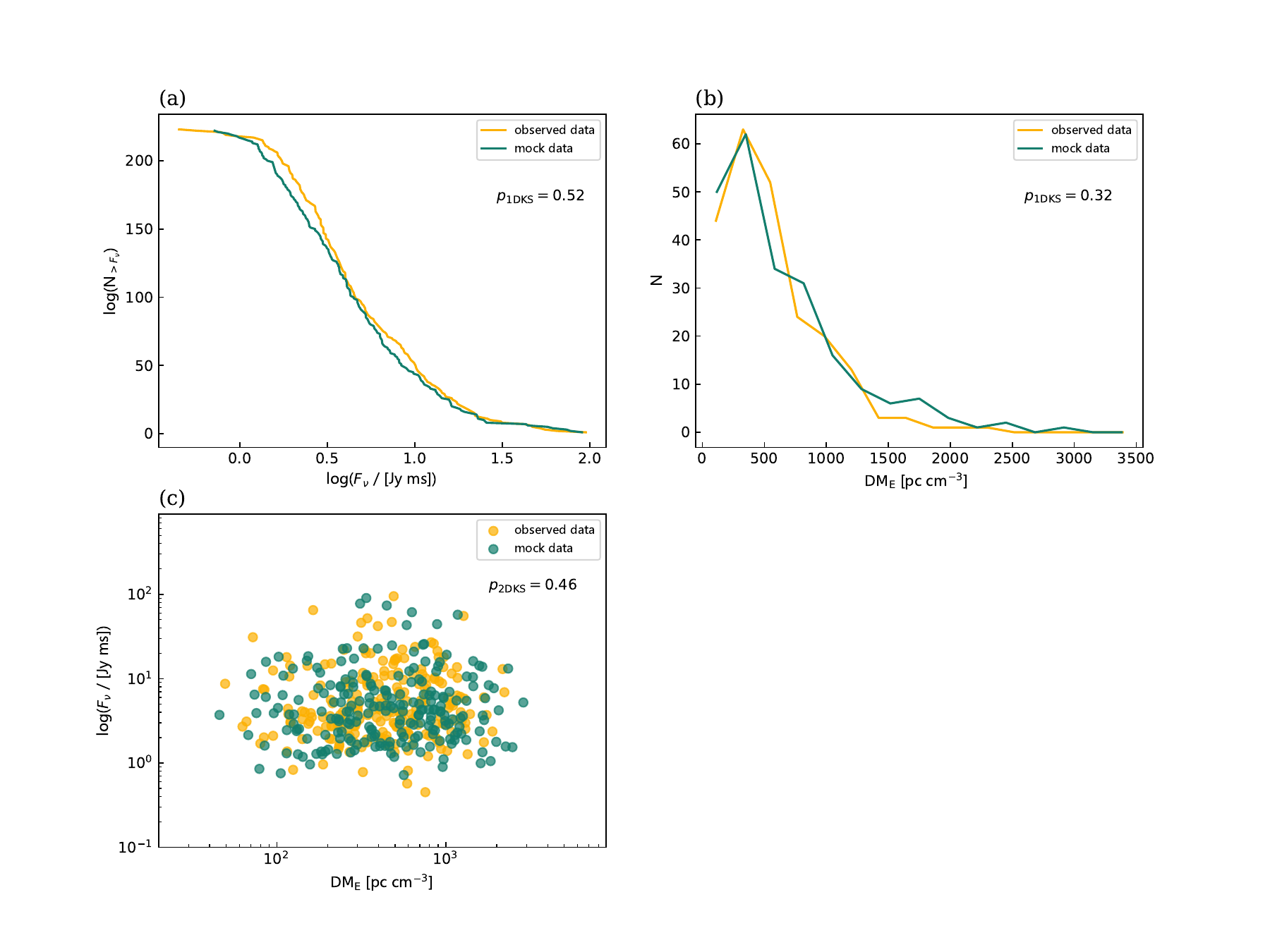}
		\caption{KS tests of the SFH model versus the $F_{\nu}$ and $\rm DM_E$ of the observed sample.  (a) 1D KS test versus the distribution of ${\rm{log}}~F_{\nu}$. (b) 1D KS test versus the distribution of $\rm DM_E$. (c) 2D KS test versus the  $\rm DM_E$ - ${\rm{log}}~F_{\nu}$ distribution.}
		\label{fig:comp1}
	\end{figure}

	\begin{figure}
		\centering
		\includegraphics[width=1\hsize]{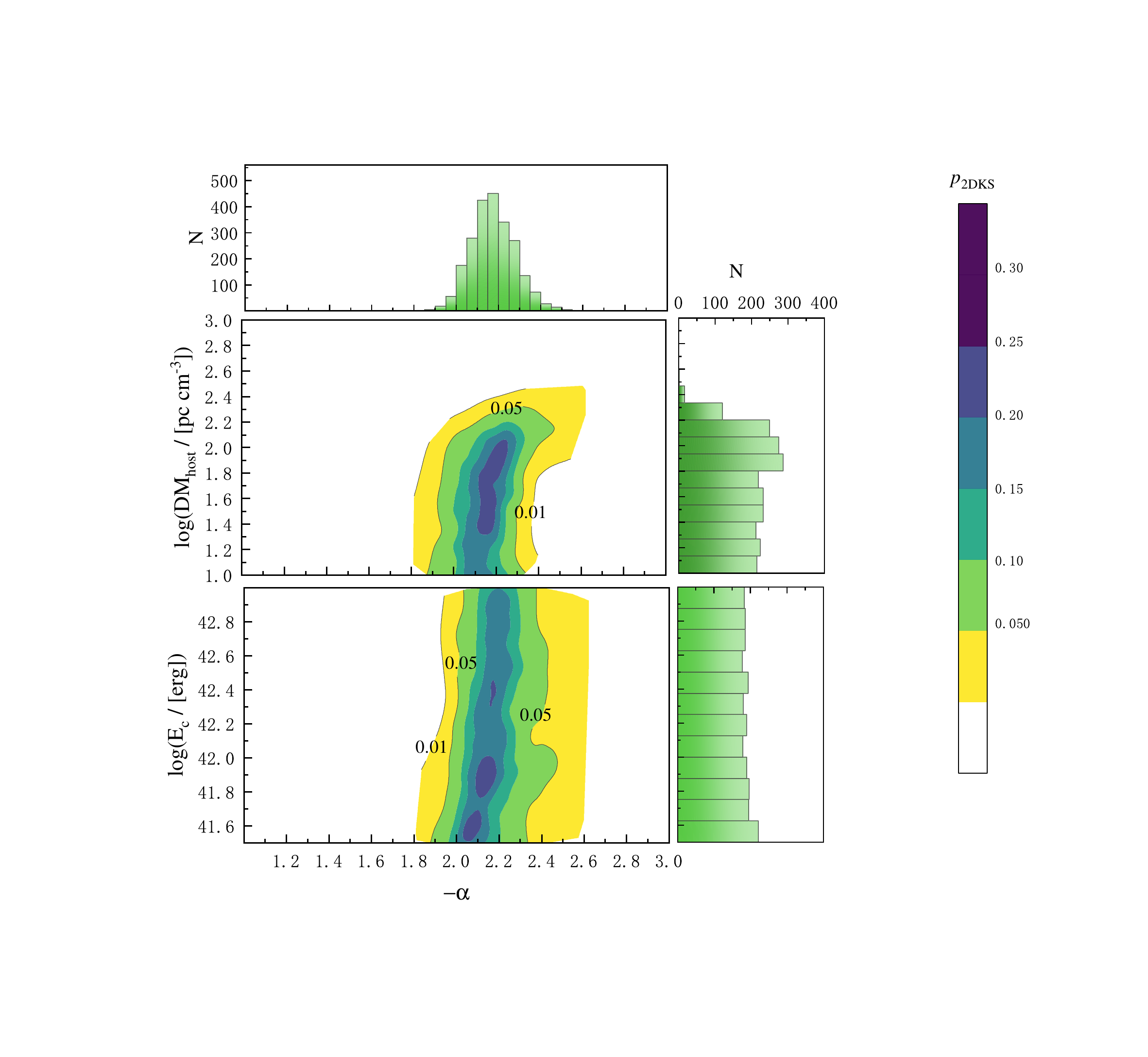} 
		\caption{Contours of $p_{\rm 2DKS} $ in the planes of $\alpha - E_{c}$ and $\rm \alpha - log(\overline{DM}_{host})$ for the SFH model. The color regions indicate parameters with $p_{\rm 2DKS}  \geqslant $  0.01, with the color intensity representing the value of $p_{\rm 2DKS} $. The contour lines corresponding to $p_{\rm 2DKS} $ = 0.01 and 0.05 are marked. The marginalized 1D distribution of each parameter with $p_{\rm 2DKS} \geqslant 0.05$ is also plotted as histograms.
		}
		\label{fig:sfr}
	\end{figure}

	\begin{figure}[ht!]
	\centering
	\includegraphics[width=1\hsize]{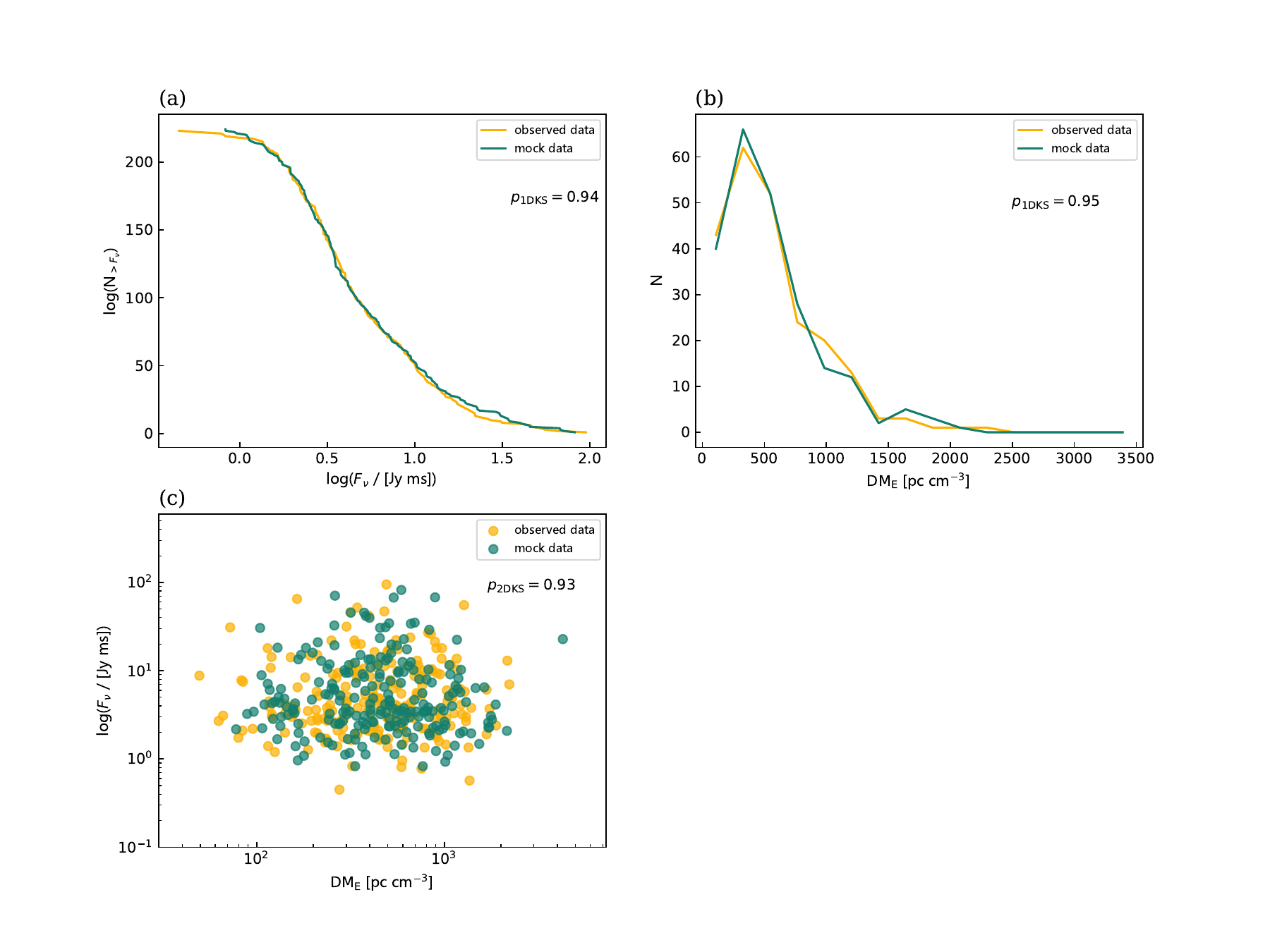}
	\caption{KS tests of the delayed SFH model against the data with $\bar{\tau}=2.2$ Gyr (similar to Fig. 2).  $\alpha=-1.7$ and $E_c=10^{41.7}$ erg are adapted for the energy function, and  $\rm \overline{DM}_{host} = 56 ~pc~cm^{-3}$ for the distribution of the host DM contribution.}
	\label{fig:comp2}
\end{figure}
	
	\begin{figure}
		\centering
		\includegraphics[width=1\hsize]{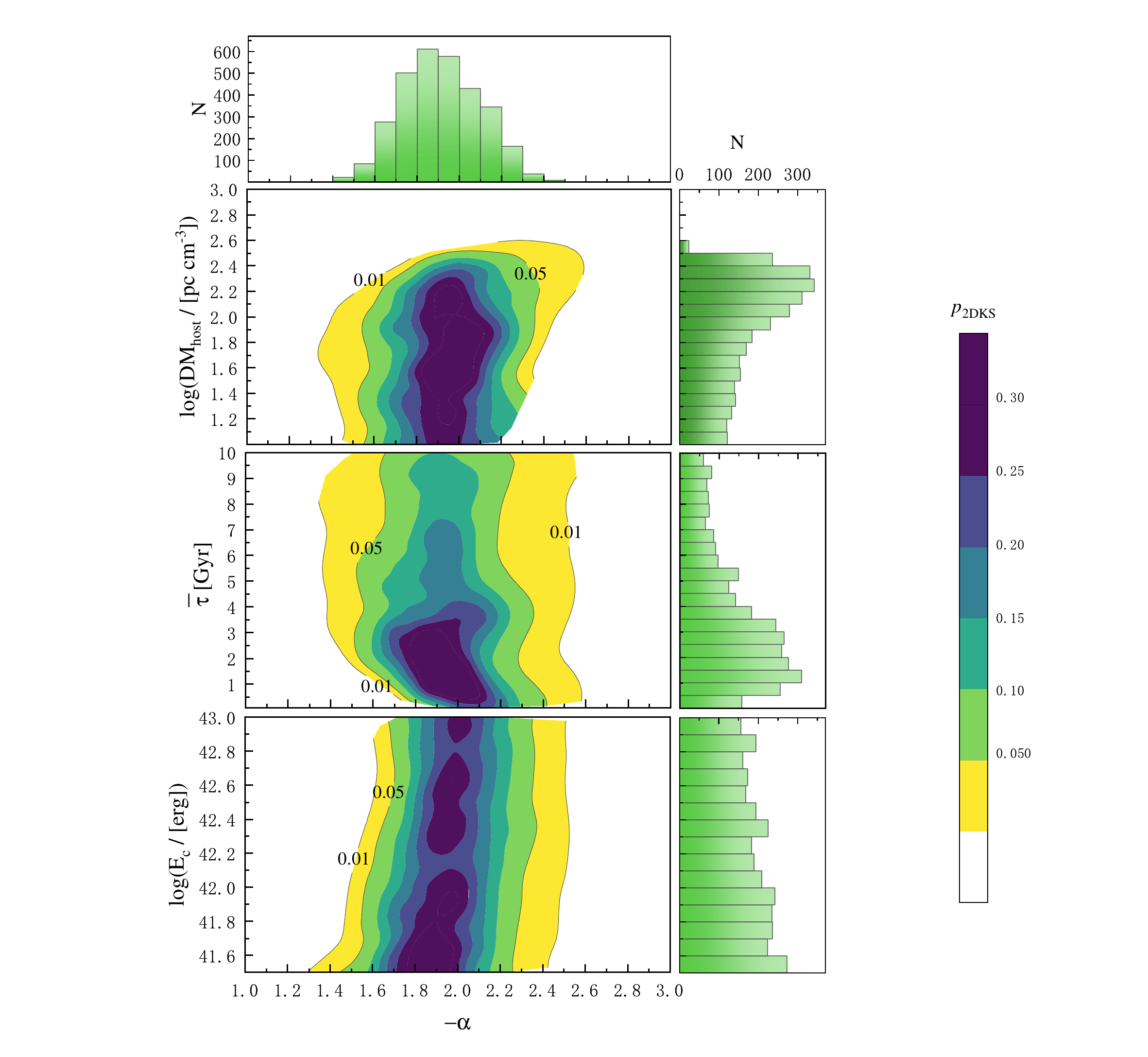}
		\caption{Contours of $p_{\rm 2DKS} $ in the planes of $\alpha - E_{c}$, $\alpha - \bar{\tau}$, and $\rm \alpha - log(\overline{DM}_{host})$. The colored regions within the contours represent parameter spaces where $p_{\rm 2DKS} \geqslant $  0.01. The contour lines corresponding to $p_{\rm 2DKS} $ = 0.01 and 0.05 are marked. The marginalized 1D distribution of the each parameter with $p_{\rm 2DKS} \geqslant 0.05$ is also  plotted as histograms.}
		\label{fig:contour2}
	\end{figure}

	{We investigated the delayed SFH model using the same method. Since the SFH model is not rejected by the data, the  delayed SFH  model should also be  not rejected by the data  in some parameter spaces because  the delay model is a generalization of the SFH model. Surprisingly, we find the delay model is not rejected by the data over a wide range of parameter spaces. Specifically, a comparison between the mock data and the observed one is  shown in figure \ref{fig:comp2},  with $p_{\rm 2DKS}=0.92$ of the 2D KS test. Also, we selected 5000 parameter combinations with $p_{\rm 2DKS} \geqslant 0.01 $  and plotted their contour map in figure \ref{fig:contour2}, the marginalized distribution of each parameter with $p_{\rm 2DKS} \geqslant 0.05$ also shown.  As one can see, the delay model with any time delay cannot be rejected by data, also shown in table \ref{tab:delay}. Here, again, $E_c$ remains unconstrained by the data. Similarly, we infer $ \overline{\rm DM}_{\rm host} <363 ~ \rm pc ~cm^{-3} $  and  $ \alpha=-1.8^{+0.4}_{-0.7} $ at the 95\% confidence level, which perfectly encompasses the Accepted range of the SFH model.  
	Although we do not rule out the SFH model,  we find that, interestingly, the delayed model more easily generates  mock samples that match the observed data. Moreover, the p-values from the KS test for the delayed model are generally significantly larger, indicating a better fit to the observed FRB sample compared to the SFH model, as one can see in Figs. \ref{fig:comp2} and \ref{fig:contour2}.  While the data do not provide a precise constraint on the delay time, the most suitable  delay time appears to be $\bar{\tau} \sim 1$  Gyr, according to the marginalized 1D distribution of $\bar{\tau}$.  	}

	\begin{table}
		\centering
		\caption{ Accepted range of free parameters for SFH model. }
		\renewcommand{\arraystretch}{1.3} 
		\begin{tabular}{cccccccccc}
			\hline\hline			
			Free parameters  & Prior & Accepted range\\
			\hline
			$\alpha$ & $[-3,-1]$& $[-2.5, -1.4] $\\
			${\rm log}(E_{\rm c}/ \rm erg)$ &$ [41.5,43]$& [41.5, 43]\\	
			$\bar{\tau}/ \rm Gyr $	&[0.01, 10]&[0.01, 10]\\
			$\rm {\overline{\rm DM}_{host}/pc~ cm^{-3}} $& $[10,1000]$&[10, 363]\\
			\hline
		\end{tabular}
		\label{tab:delay}
	\end{table}

	\section{Discussion}
	  {Our results show that neither the SFH model nor the delay model are ruled out by the data, which contrasts with the conclusions made by \cite{2022ApJ...924L..14Z} and \cite{2022JCAP...01..040Q}. However, it is important to note that the details of our simulations, especially the treatment of instrumental selection effects, differ from those used in these studies.}

	\medskip\noindent\textnormal{\textit{Redshift distribution models}.}
	{Neither the SFH model nor the delayed SFH model are ruled out by the data, which is consistent with the findings of \cite{2023ApJ...944..105S}. However, an important distinction between our results and theirs is that, while we do not rule out the SFH model, we observe that the data tend to prefer the delayed SFH model with an appropriate delay time. }
	
    {As mentioned in the introduction, several studies based on the CHIME/FRB sample, including \cite{2022ApJ...924L..14Z}, \cite{2022JCAP...01..040Q}, \cite{2022MNRAS.511.1961H}, \cite{2023A&A...675A..66Z}, \cite{2024ApJ...962...73L}, \cite{2024ApJ...969..123L}, \cite{2024ApJ...973L..54C}, and \cite{2024arXiv240600476Z}, have all  concluded that the delay model is needed to explain the data, explicitly rejecting the SFH model.  We believe that the discrepancy in the conclusions is due to the following.
   	Firstly, it is important to note that the  methods for handling selection effects they used are relatively primitive. 
   	In addition, they have incorrectly used the fluences (which actually are lower limits) provided in the CHIME/FRB catalog as true values, which might also have led to biases in their results. }
	
	 \medskip\noindent\textnormal{\textit{Energy function}.}
	{We find that the power-law index of the  energy function is constrained to $\alpha =-1.8^{+0.4}_{-0.7}$ regardless of the redshift distribution models of FRB population. The result is consistent with the results of $\alpha=-1.3^{+0.7}_{-0.4}$ in \cite{2023ApJ...944..105S}  within the error bars, who also used the CHIME/FRB sample but employed different approaches. It also is in broad agreement with the results obtained by \cite{2022MNRAS.511.1961H}, \cite{2024ApJ...962...73L}, and \cite{2024ApJ...969..123L}, which suggests that the energy function is not greatly affected by the models of redshift distribution  and the selection effects.
		\cite{2022MNRAS.510L..18J}  obtained a steeper index $\alpha =-2.09^{+0.14}_{-0.10}$  using the FRB sample from {ASKAP} and Parkes.}
	
	\medskip\noindent\textnormal{\textit{Host galaxy DM distribution}.}
	{We modeled the rest-frame host galaxy DM ($\rm DM_{host}$) with a log-normal distribution. { The mean value of the $\rm {DM}_{host}$ distribution is constrained to  $< 363 ~\rm ~pc~cm^{-3}$.} This result is broad agreement with  the results of \cite{2022MNRAS.510L..18J} and \cite{2023ApJ...944..105S}, who find $\rm \overline{DM}_{host} =129^{+66}_{-48} ~pc~cm^{-3}$ and $\rm \overline{DM}_{host} =84^{+69}_{-49} ~pc~cm^{-3}$, respectively.  Meanwhile, these results indicate that the host galaxy DM for the majority of FRBs is likely not large, and the cases with extremely large $\rm DM_{host}$ like FRB 190520B should be considered special and rare \citep{2022Natur.606..873N}.}
	
	\medskip\noindent\textnormal{\textit{Delay time with respect to the SFH}.}
	{Our results indicate that the redshift distribution of FRBs does not require a delayed time \(\tau\) with respect to the SFH to be consistent with the data, as the SFH model is not rejected. However, interestingly, the delay model appears to better reproduce the observed data, as evidenced by the generally higher \(p\)-values from the KS tests. While the required delay time does not have a precise constraint from our analysis, the data suggest that a delay time of $ \sim 1$ Gyr is the most likely scenario. }

	\medskip\noindent\textnormal{\textit{Event rate of FRB sources}.}
	In our simulations, it is easy to calculate the event rate of the FRBs. In order to obtain a robust estimation of the mean value of the event rate and its standard deviation, we conducted 1000 successful simulations ($ p_{\rm 2DKS}>0.05$) using randomly combined model parameters and performed statistical analysis. To get 223 detectable FRBs for a single mock sample,  one needs to generate a number of $4 \times 10^{4}$ FRBs on average.  
	By combining the corresponding survey duration ($\Delta t=214.8$
	days) and the field of view of CHIME/FRB ($\sim$ 0.3\% of the sky;  \citealt{2021ApJS..257...59C}), one can calculate the bursts rate of FRBs. 

	If the fluence $ F_{\nu}$ is used as a reference point, then the all-sky rate is $\dot{R}_{\rm  sky}(F_{\nu} \rm >5~Jy~ ms)=216^{+21}_{-19} ~ sky^{-1}~day^{-1}$, which is smaller than the value inferred in \cite{2023ApJS..264...53C} by a factor of 2,  but still within the error bars  if the system uncertainty is considered.  If one counts the bursts within 1 Gpc, the local event rate density of FRBs can be obtained $\dot{\rho}_0(E>10^{38}~ \rm erg)=2.3^{+2.4}_{-1.2} \times 10^5 ~Gpc^{-3}~yr^{-1}$. Interestingly, \cite{2019NatAs...3..928R} also found a similarly high volumetric rate for the FRBs by assuming $\rm DM_{host}=50~pc~cm^{-3}$. Such a high volumetric rate is a serious challenge to the progenitor models of FRBs. If a pivot energy of $10^{39}$ erg is applied, one has $\dot{\rho}_0(E>10^{39} ~\rm erg)= 3.9^{+2.5}_{-1.2} \times 10^4 ~Gpc^{-3}~yr^{-1}$, which is consistent with the results of \cite{2023ApJ...944..105S}. For more information on the event rate, please refer to the table \ref{tab:De}  and table \ref{tab:De1}. The errors for the event rate are all at 1 $\sigma$ confidence level.
	{As one can see, the local event rate density is strongly dependent on the reference energy.  As shown in table \ref{tab:De}, a lower reference energy results in a higher event rate density, which is roughly consistent with the scaling relation of $\dot{\rho}_0(>E) \propto E^{\alpha+1}$.}

	\begin{table*}
		\centering
		\caption{The Local rate density above a pivot energy. }
		\renewcommand{\arraystretch}{1.5} 
		\begin{tabular}{cccccccccc}
			\hline\hline
			
			Rates  &~&  $\rm >10^{38} erg$& $\rm >10^{39} erg$& $\rm >10^{40} erg$& $\rm >10^{41} erg$\\
			\hline		
			Local rate density ($\rm Gpc^{-3}~yr^{-1}$) &~& $2.3^{+2.4}_{-1.2}  \times 10^5$& $3.9^{+2.5}_{-1.2} \times 10^4$& $5.5^{+3.0}_{-1.9} \times 10^3$& $8.4^{+5.7}_{-3.4} \times 10^2$ \\
			\hline
			
		\end{tabular}
		\label{tab:De}
	\end{table*}
	\indent
	\begin{table*}
		\centering
		\caption{All-sky rate above a pivot fluence.}
		\renewcommand{\arraystretch}{1.5}
		\begin{tabular}{cccccccccc}
			\hline\hline
			
			Rates  &~& $\rm >0.1~Jy~ms$ & $\rm >0.5~Jy~ms$& $\rm >1~Jy~ms$& $\rm >5~Jy~ms$& $\rm >10~Jy~ms$& $\rm >50~Jy~ms$& $\rm >100~Jy~ms$\\
			\hline
			All-sky rate ($\rm sky^{-1} day^{-1}$)&~& $4453^{+3146}_{-1843} $ & $1469^{+447}_{-343} $& $865^{+141}_{-121} $ & $216^{+21}_{-19} $& $110^{+21}_{-18} $& $18^{+12}_{-7} $& $8^{+6}_{-3} $\\
			\hline
			
		\end{tabular}
		\label{tab:De1}
	\end{table*}

	\section{Conclusions}  \label{sec:Conclusions} 
	{In this work we conducted extensive Monte Carlo simulations to test the redshift distribution models of CHIME FRBs. Given the complexity of selection effects in the CHIME/FRB catalog, we carefully modeled these effects to ensure the robustness of our analysis. For the redshift distribution, we examined two key models inspired by current FRB theories: the cosmic formation rate of FRBs is proportional to the SFH, and it is delayed with respect to the SFH. By systematically exploring the complete parameter spaces for these models, we identified configurations that are consistent with observational data.  Our main findings are as follows: }
	
	{$\rm (i)$ Within a specific range of model parameters, the SFH model cannot be ruled out by the data, although the data tend to favor a small delay. 
	Our results from testing the redshift distribution of FRBs are generally consistent with the results of  \cite{2023ApJ...944..105S}. }
 
	{$\rm (ii)$ The CHIME FRB sources have a very high local rate density of $\rm \dot{\rho}_0 (>10^{38} \rm erg) =2.3^{+2.4}_{-1.2} \times 10^5 ~Gpc^{-3}~yr^{-1}$.  The estimated local rate density in this work aligns well with the results of \cite{2019NatAs...3..928R} and \cite{2023ApJ...944..105S}, and the estimation for the all-sky rate is close to the results of \cite{ 2023ApJS..264...53C}.}
	
	{Given that the hypothesis that the FRB population tracks the SFH cannot be ruled out, in accordance with Occam's razor, our results support the idea that FRBs  primarily originate from young stellar populations, such as magnetars formed through core-collapse supernovae. Nevertheless, the extremely high volumetric rate of FRBs imposes stringent constraints on the progenitor models. For core-collapse supernovae, which could occur at a rate of approximately $10^5~\rm Gpc^{-3}~yr^{-1}$ \citep{2012ApJ...759..107K}. However, it is uncertain what fraction of these events produce the compact objects capable of driving FRBs. Notably, roughly half of these products are expected to be magnetars \citep{2008MNRAS..391..2009K}, which are widely accepted as the most plausible candidates for FRB central engines. Thus, the core-collapse supernova channel  alone is  insufficient  to account for the majority of FRB sources.}
	In light of these challenges, it has been argued that if FRBs are inherently repetitive, this may explain the high burst rate, as a single FRB source could produce multiple bursts  \citep{2018ApJ...858...89C,2019NatAs...3..928R}. 
	However, in the sample used in this work, we only counted the first detected burst from each repeating source, and thus each FRB corresponds to an independent source.  
	Furthermore, in our simulation, each mock FRB corresponds to an independent source, and we do not allow multiple events from the same source in the Monte Carlo process. Therefore, the event rate we obtain directly reflects the rate of the FRB sources, not the burst rate. 
	However, it is important to clarify that the event rate we derived in this work should actually represent the rate of the FRB sources rather than  the rate of the bursts\footnote{This is because in the sample used in this work, we only count the first detected burst from each repeating source, and thus each FRB corresponds to an independent source.  Furthermore, in our simulation, each mock FRB corresponds to an independent source, and we do not allow multiple events from the same source in the Monte Carlo process. Therefore, the event rate we obtain directly reflects the rate of the FRB sources, not the burst rate.}. Therefore, the repetition of bursts from a single source cannot explain the high rate of the sources  itself.

	{We caution that our conclusions are drawn based on the current CHIME/FRB sample and, thus, are inherently subject to its limitations. Given the high event rate of FRBs, it is foreseeable that significantly larger samples will become available in the near future. An expanded dataset would enable further studies that could potentially strengthen our conclusions or force us to revise them. Moreover, effectively correcting for CHIME's complex selection effects remains a critical aspect of such research. The beam model used in this study is an early version, and future releases of more refined versions are expected to enhance the accuracy of such analyses.
	Only with a large enough sample and a clearer understanding of CHIME's selection effects can robust conclusions be drawn. We anticipate that future advancements in observations and methodologies will provide stronger constraints on the progenitor model of FRBs, improving our understanding of their origins.}

    {During the review process of this manuscript, we came across an independent study by \cite{2024A&A...690A.377W}. In their study, they employed the open-source Python package frbpoppy to perform a comprehensive analysis of the characteristics of the FRB population based on CHIME/FRB Catalog 1. They reached a conclusion consistent with ours: the hypothesis that the FRB population follows the SFH cannot be rule out.}
	
\begin{acknowledgements}
	This work is supported by the National Natural Science Foundation of China (grant No. 12203013) and the Guangxi Science Foundation (grant Nos. AD22035171 and 2023GXNSFBA026030).  We are also grateful to the referee for very useful comments on the manuscript.
\end{acknowledgements}

	\bibliographystyle{aa}
	\bibliography{ref}
	
\end{document}